# HLC2: a highly efficient cross-matching framework for large astronomical catalogues on heterogeneous computing environments

Yajie Zhang,[1,2] Ce Yu,[1,2]★ Chao Sun,[1,2] Jian Xiao,[1,2] Kun Li,[1,2] Yifei Mu[1,2] and Chenzhou Cui[2,3]

[1]*College of Intelligence and Computing, Tianjin University, No. 135 Yaguan Road, Haihe Education Park, Tianjin 300350, China*
[2]*Technical R&D Innovation Center, National Astronomical Data Center, No.135 Yaguan Road, Haihe Education Park, Tianjin 300350, China*
[3]*National Astronomical Observatories, Chinese Academy of Sciences, No.20 Datun Road, Chaoyang District, Beijing 100012, China*

**ABSTRACT**
Cross-matching operation, which is to find corresponding data for the same celestial object or region from multiple catalogues, is indispensable to astronomical data analysis and research. Due to the large amount of astronomical catalogues generated by the ongoing and next-generation large-scale sky surveys, the time complexity of the cross-matching is increasing dramatically. Heterogeneous computing environments provide a theoretical possibility to accelerate the cross-matching, but the performance advantages of heterogeneous computing resources have not been fully utilized. To meet the challenge of cross-matching for substantial increasing amount of astronomical observation data, this paper proposes **H**eterogeneous-computing-enabled **L**arge **C**atalogue **C**ross-matcher (HLC2), a high-performance cross-matching framework based on spherical position deviation on CPU-GPU heterogeneous computing platforms. It supports scalable and flexible cross-matching and can be directly applied to the fusion of large astronomical catalogues from survey missions and astronomical data centres. A performance estimation model is proposed to locate the performance bottlenecks and guide the optimizations. A two-level partitioning strategy is designed to generate an optimized data placement according to the positions of celestial objects to increase throughput. To make HLC2 a more adaptive solution, the architecture-aware task splitting, thread parallelization, and concurrent scheduling strategies are designed and integrated. Moreover, a novel quad-direction strategy is proposed for the boundary problem to effectively balance performance and completeness. We have experimentally evaluated HLC2 using public released catalogue data. Experiments demonstrate that HLC2 scales well on different sizes of catalogues and the cross-matching speed is significantly improved compared to the state-of-the-art cross-matchers.

**Key words:** methods: data analysis – techniques: miscellaneous – catalogues – software: public release.

## 1 INTRODUCTION

In astronomy research, cross-matching is a crucial operation to find corresponding data for the same object or region from multiple sources. The processing of cross-matching is essentially a spatial matching operation, which is highly data-parallel. It is the prerequisite for a fusion of observation data from different bands in multiband astronomy and acquisition of information about the same celestial object at different times in time-domain astronomy. Efficient execution of cross-matching is not only beneficial for data fusion from different wavelengths or telescopes, but also necessary for time-series data reconstruction, transient candidate discovery, and subsequent survey deployment. With the increase of astronomical observation equipment and the improvement of observation ability, astronomers make efforts to achieve full temporal and spatial monitoring of the order of hours or higher frequency. Therefore, ongoing sky survey projects have produced astronomical catalogues of a vast number of objects, such as SDSS (SDSS 2015) and *Gaia* (Gaia et al. 2018), each containing more than 1 billion objects (Jia, Luo & Fan 2016). A single observation image can parse out nearly a million catalogue records (GAIA 2020). In big data era, the volume of astronomical data continues to grow and the National Academy of Sciences notes that some observatories will bring 500 PB of data by the end of 2030 across all missions (National Academies of Sciences et al. 2021). It brings a challenge for astronomers to cross-match and extract information from these data. Dealing with every source pairwise is impractical comparison due to high computational complexity.

For the problem of cross-matching between two large-scale catalogues, index partitioning is an efficient way to reduce the unnecessary computation. Previous works have proposed basic spatial indices, including the hierarchical equal area isoLatitude pixelation (HEALPix; Gorski et al. 2005), hierarchical triangular mesh (HTM; Kunszt, Szalay & Thakar 2001), Zones (Gray et al. 2004; Gray, Nieto-Santisteban & Szalay 2007), etc. These schemes can reduce part of the computation but are accompanied with the emergence of boundary problems. Solutions to address the problems mainly include adding redundant data on the boundary (Zhao et al. 2009), using dual indices (Du et al. 2014; Yu et al. 2020), etc. Facing with data avalanche and high decentralization, another way of improving efficiency is to apply data parallelization in various ways. So far, only a few studies have been implemented on multicore CPU (Nieto-santisteban, Szalay & Thakar 2005; Zhao et al. 2009; Du et al. 2014; Riccio et al. 2017; Soumagnac & Ofek 2018) and GPUs (Budavari & Lee 2013; Jia et al. 2016; Jia & Luo 2016).

★ E-mail: yuce@tju.edu.cn

Although aforementioned methods have done a lot on improving the overall performance, problems, and technical challenges still remain. For the problem of cross-matching between two catalogues with more than millions of sources, the performance needs to be further accelerated. Meanwhile, when heterogeneous computing resources are used for cross-matching processing, efficient partitioning and parallel organization strategies are required to take full advantage of heterogeneous processors. In addition, as the number of celestial records in the catalogues increases, the existing strategy loses part of the matching completeness for the speed improvement and the generation and processing of redundant data for boundary problems make the calculation more time-consuming. So, the balance of speed and completeness can be further explored.

To deal with the challenges mentioned above, we propose and implement an efficient cross-matching framework named HLC2 on CPU-GPU heterogeneous computing environments to speed up the cross-matching processing. The key features of HLC2 can be summarized as follows:

(i) A flexible cross-matching framework. HLC2 can efficiently deal with the cross-matching calculation of catalogues with hundred millions of sources. It scales nicely for various NVIDIA GPU architectures and sizes of astronomical catalogues, and the modular construction can also ensure the scalability.

(ii) A new solution to boundary problems. The quad-direction strategy is designed as an efficient solution to balance the loss of computational completeness while improving the performance of the cross-matching. Experimental results show improved completeness with similar speed to the next best approach.

(iii) High-performance parallel processing. A performance prediction model is proposed for theoretical analysis on the influencing factors of performance. Based on the profiling, optimization strategies are proposed and implemented to achieve efficient CPU-GPU co-processing. A two-level partitioning strategy is proposed on the CPU, while dynamic thread organization and I/O optimization are conducted on the GPU. The speed of cross-matching is considerably improved compared to the state-of-the-art cross-matchers.

The rest of the paper is organized as follows. Section 2 briefly introduces the details of cross-matching and presents the existing work related to astronomical pseudo-spherical indexing schemes and parallel acceleration methods. Sections 3 and 4 focus on the design of our HLC2 framework and optimization strategies. Results of related experiments are described in Section 5. Section 6 contains our conclusion and discusses future work.

## 2 BACKGROUND AND RELATED WORK

### 2.1 Cross-matching of astronomical catalogues

A catalogue is a 2D data table containing celestial information such as right ascension, declination, magnitude, etc. Due to different data acquisition, calibration methods, and observation instruments, the same object may have slightly different coordinates in different catalogues (Nieto-santisteban et al. 2005). Consequently, a distance threshold based on calibration error is the condition to judge whether two celestial objects are the same or not. In general, the angular distance between two objects $O_1$ and $O_2$ can be calculated according to the following rules (Zhang & Zhao 2003).

(1) Distance formula:
The coordinates of two objects $O_1$ and $O_2$ are represented by $(\alpha_1, \delta_1)$ and $(\alpha_2, \delta_2)$. According to the law of spherical cosines, the spherical angle distance $d$ between A and B can be expressed as:

$$d = \arccos\left(\sin \delta_1 \sin \delta_2 + \cos \delta_1 \cos \delta_2 \cos(\alpha_1 - \alpha_2)\right). \quad (1)$$

When the angular distance between $O_1$ and $O_2$ is very small,

$$\delta \approx (\delta_1 + \delta_2)/2, \quad (2)$$

so the spherical angle distance can be calculated as

$$d^2 = ((\alpha_1 - \alpha_2) \times \cos \delta)^2 + (\delta_1 - \delta_2)^2. \quad (3)$$

(2) Formula for successful cross-matching:

$$d = \sqrt{((\alpha_1 - \alpha_2) \times \cos \delta)^2 + (\delta_1 - \delta_2)^2} \leq 3\sqrt{r_1^2 + r_2^2}, \quad (4)$$

where $r_1$ and $r_2$ are the error radius of the two catalogues. When the distance $d$ between the two points satisfies the inequality, the matching can be considered successful, which means the records represent the information of the same celestial object. So, for two catalogue data sets of size $m$ and $n$, respectively, the time complexity is $O(mn)$, which is unacceptable for large data volume generated by recent large-scale sky surveys.

### 2.2 Astronomical indexing schemes

With respect to massive data processing, reducing unnecessary comparisons by indexing is an essential means to improve efficiency. Since simple nested loop matching for each directory requires massive unnecessary pairwise comparisons, previous work has proposed various spatial indices. Three mainstream indexing methods are HEALPix (Gorski et al. 2005), HTM (Kunszt et al. 2001), and Zones (Gray et al. 2004; Gray et al. 2007), which partition the sphere into diamond-shaped, triangle, and rectangle cells, respectively. The common idea behind them is to divide the sky into a fixed number of regions for a given resolution, with each object associated with an integer index of the region containing that object. Celestial objects that are close to each other are assigned to the same index or index of nearby regions. HEALPix organizes indices in a hierarchy of resolutions with the entire sky as the root node and regions at the finest resolution as leaves. Newly created regions are appended a label in a finite collection to the index of its parent. Subdivided regions under the same parent region are named in a nested clockwise fashion. Researchers proposed quad tree cube (Q3C), which is partitioned into a cube (Koposov & Bartunov 2006). Landais, Ochsenbein & Simon (2013) were inspired from Q3C to build the 2D PostgreSQL library HEALPix-Tree-C (H3C), the functionality of which is same as Q3C, but works with HEALPix.

In addition to optimization by spatial index partitioning, another way is by creating a search tree. The difference is that tree indexing can reduce the computational complexity from $O(n^2)$ to $O(n \log n)$, while spatial indexing prunes the records with different indices in the two catalogues and does not compute them directly. Zhao's previous work compared the two optimization methods of grid indexing for sky regions and single dimensional B-tree searching for declination in the same experimental environment (Zhao et al. 2009). Compared with the parallel algorithm of declination single dimensional index, the performance of grid indexing is improved by about 60 times. More recently, Li et al. (2019) proposed a multiband cross-matching schema with a specially designed catalogue format and data layout based on KD-tree, which can support faster query response than Q3C and H3C after sources scale up to 100 million. Tree structure has shown considerable usefulness on the retrieval problems.

The aforementioned mainstream spatial indices reduce unnecessary comparisons, but they have also brought along with a negative effect that two different records of the same object may be divided into different adjacent regions. As a result, the two records will not be compared in the subsequent calculation based on spherical position deviation, so the corresponding data of the object will be lost. Then, a part of astronomical discoveries will be missed. In order to simplify computation procedure and improve performance, many cross-matching methods ignore the boundary problem, which results in a certain degree of result loss. Meanwhile, extra processing strategies of edge data have also existed when conducting the index-based partition. Zhao et al. (2009) came up with a remedy of increasing the boundary data by a quick bit-operation algorithm. However, it brings along with too much redundant data, which greatly affects the real-time performance on large data sets. Du et al. (2014) proposed a dual indexing method that combined HEALPix and HTM reduced part of the data leakage, which was also used in time-series data reconstruction (Yu et al. 2020). Even though tremendous improvements have been achieved, we note that the increase of computational load brought by these methods is not conducive to computational efficiency. Therefore, for the problem of cross-matching between two catalogues with more than millions of sources, an efficient edge solving algorithm is needed to maintain the balance of computational complexity and completeness.

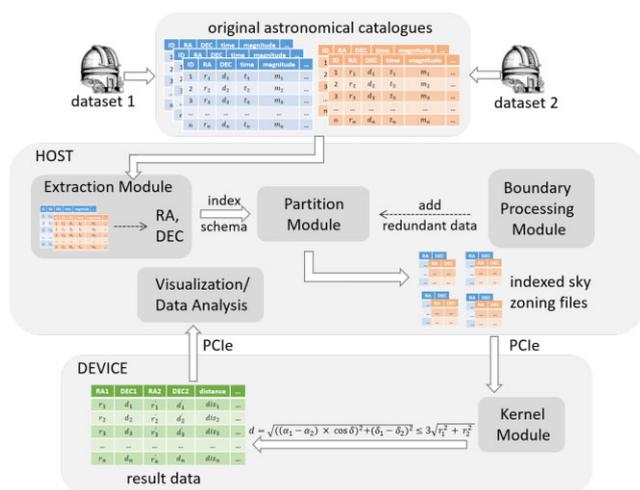

**Figure 1.** Overview of the basic HLC2 architecture. It mainly consists of four modules, including extraction, which extracts position information involved in the cross-matching calculation; partition, which divides the catalogues for parallel computation; boundary processing for the balance of completeness and performance; finally kernel for inter-catalogue parallelization on the GPU.

### 2.3 Existing astronomical catalogue cross-matching accelerations

Modern parallel processors, such as GPU, have been widely adopted in various applications to improve the processing performance. Besides researchers represented by Zhao et al. (2009) proposed acceleration methods using MPI, Du et al. (2014) used the thread pool technique to speed up the cross-matching. Jia et al. (2016) explored cross-matching strategies on CPU-GPU clusters. In addition, mcatCS (Li et al. 2019) is also a distributed cross-matching scheme to efficiently integrate celestial object data. However, it focuses on single record cross-matching with low complexity and is therefore time-consuming for large-scale catalogue matching. Meanwhile, Wang et al. (2013) proposed two parallel algorithms based on cone search that were implemented on GPU, and Budavari & Lee (2013) applied it on multiple GPUs. The aforementioned algorithms handle cross-matching of millions of catalogue data in several seconds. However, further improvements of performance and resource utilization are needed, and cross-matching computations with larger data volumes need to be solved.

There are also other acceleration methods based on Hadoop (Li et al. 2014) and Spark framework (Jia & Luo 2016; Zečević et al. 2019). Additionally, several cross-matching tools have been developed to handle massive catalogues, represented by TOPCAT (Taylor 2011), SIMBAD (Wenger et al. 2000), VizieR (Ochsenbein, Bauer & Marcout 2000), CDS-Xmatch (Boch, Pineau & Derriere 2014) and NASA/IPAC (Helou et al. 1991). These offer querying of observation data, some editing, and analyzing functions. In recent years, other web applications are developed like catsHTM (Soumagnac & Ofek 2018). Nevertheless, the current number of jobs and users may be limited because in web-base applications, cross-matching calculations account for the majority of server consumption. Meanwhile, a stand-alone command-line PYTHON tool C3 is proposed (Riccio et al. 2017), having comparable performance compared with CDS-Xmatch and TOPCAT.

## 3 HETEROGENEOUS-COMPUTING-ENABLED LARGE CATALOGUE CROSS-MATCHER

This work focuses on an accelerated cross-matching method for large-scale astronomical catalogues in the way of CPU-GPU co-processing. The cross-matching computation based on spherical position deviation is migrated to CPU-GPU heterogeneous computing environments. A general cross-matching framework named HLC2 is proposed, which is suitable for efficient astronomical catalogue processing.

### 3.1 Basic architectural design of HLC2

Fig. 1 shows the basic design of HLC2. Our algorithm is encapsulated into an end-to-end command-line-based cross-matching tool running on the Linux platform, which is implemented in C and C++, and CUDA for NVIDIA GPUs. GPUs typically have more computing cores than CPUs in the form of streaming processors, which can be further organized into groups stream multiprocessors (SMs). The SIMT fashion of GPU programs is beneficial for various astronomical information processing tasks, especially for the angular distance calculation of the cross-matching problem, because each pair of catalogue records follows the same calculation mode and has no dependence on other data.

Meanwhile, module-based software architecture is implemented to ensure flexibility and scalability. It mainly consists of four modules, namely extraction, partition, boundary processing and kernel. The workflow for HLC2 to perform cross-matching is as follows:

On the CPU, namely host side, first HLC2 invokes **extraction module** for raw data initialization. Astronomers may input two catalogues from different survey projects in different sizes. So, it extracts celestial position information (mainly *RA*, *DEC*) from therein and filters out useless information. Next, the **partition module** divides the filtered catalogue data into different celestial regions based on the HEALPix index, which can be replaced by other schema mentioned in Section 2.2. Records from one catalogue are computed only with data from the same region of another catalogue. In the meantime, during data partitioning, the **boundary processing module** is called to maintain the completeness of the cross-matching calculation without too much redundant data added, which may affect the calculation performance. The strategies used to deal with boundary problems are detailed in Section 4.1.2. The pairs of celestial records in the indexed sky zoning files are then transferred to the GPU.

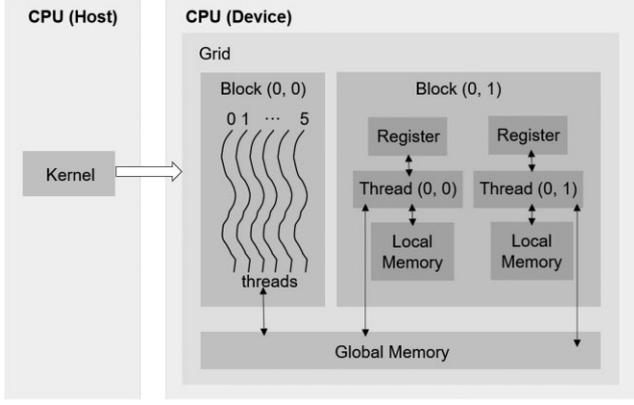

**Figure 2.** The basic hierarchy of CUDA thread and memory. Each thread has its own register and local memory, and each grid has its own global memory, the contents of which can be read and modified by threads of different thread blocks.

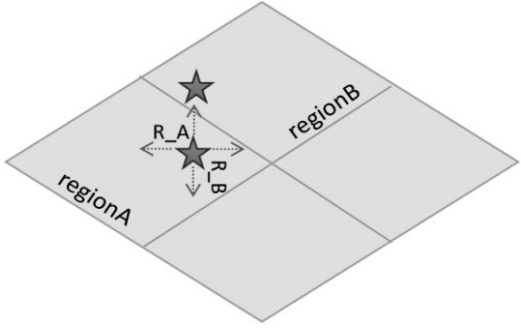

**Figure 3.** Schematic diagram of the quad-direction strategy. Solid lines in the figure represent schematic lines for the index partition. After adding and subtracting error radius of the right ascension and declination (illustrated by the dotted lines), the celestial object from *regionA* (displayed in the shape of pentagram) falls into *regionB*. So, the record of the celestial object from *regionA* will be added into *regionB*, according to our strategy.

On the GPU, namely device side, the **kernel module** adopts inter-catalogue parallelization to calculate the angular distances according to the method described in Section 2.1. As illustrated in Fig. 2, the concept of the hierarchical structure of threads is proposed by NVIDIA to facilitate thread organization (Sanders & Kandrot 2010). Since the catalogue is 1D text data, we use 1D thread hierarchy correspondingly. The computation of the distance calculations of the pairs of celestial records in the indexed sky zoning files is divided to various threads. Each thread performs the task independently with no communication. Finally, the cross-matching results are transferred to CPU and exported the final products. Results can also be visualized, which is under development.

### 3.2 Quad-direction strategy for boundary problems

The boundary problem is a main factor that affects the completeness of the matching results in large-scale catalogue processing. As shown in Fig. 3, an object that falls on the edge of a block (*regionA*) may fall outside the edge of that block in another catalogue and into its neighbour (*regionB*). In the boundary processing module, we propose an efficient method named quad-direction strategy to balance the completeness and performance. Algorithm 1 shows the implementation details of our boundary processing using quad-direction strategy. When constructing the spatial spherical index, the possible boundary objects are added into all corresponding regions by increasing the judgement of the possibility of the objects being in the adjacent index regions. The illustration of quad-direction strategy is shown in Fig. 3. For each object near the boundary, our method will judge whether the point after the error radius of the right ascension and declination ($R\_A$ and $R\_B$) is added or subtracted, respectively, from its position falls in the adjacent region. If so, the information of this object is added into the corresponding adjacent sky region. The catalogue of the celestial object in *regionA* will fall into *regionB* after $R\_B$ is added to the declination, so this catalogue is also involved in the calculation of *regionB*. Celestial records are still only involved in the distance calculation within the calculation block, and the index number, right ascension and declination remain unchanged. By this way, we expanded the scope of each data block divided depending on the granularity of the partitioning level selected by the user to reduce part of the match loss.

---

**Algorithm 1** Quad-direction strategy for boundary processing
---
**Require:** two catalogues A and B;
**Ensure:** indexed sky zones for parallel cross-matching;
1: initial map $record_A$, $record_B$;
2: **for all** $RA$, $DEC$ in $inputlines_{A/B}$ **do**
3:     $hpid \leftarrow hp\_nestid(RA, DEC, order)$
4:     $ra\_plus \leftarrow ra + R\_A$
5:     $ra\_minus \leftarrow ra - R\_A$
6:     $dec\_plus \leftarrow dec + R\_B$
7:     $dec\_minus \leftarrow dec + R\_B$
8:     $hpid_{up} \leftarrow hp\_nestid(ra\_plus, DEC, order)$
9:     $hpid_{down} \leftarrow hp\_nestid(ra\_minus, DEC, order)$
10:    $hpid_{left} \leftarrow hp\_nestid(RA, dec\_minus, order)$
11:    $hpid_{right} \leftarrow hp\_nestid(RA, dec\_plus, order)$
12:    **if** $record_{A/B}.find(hpid) = record_{A/B}.end()$ **then**
13:        $record_{A/B}.insert((hpid, line))$;
14:    **end if**
15:    **if** $hpid_{up/down/left/right} \ne hpid$ **then**
16:        **if** $record_{A/B}.find(hpid_{up/down/left/right})$
17:            $= record_{A/B}.end()$ **then**
18:            $record_{A/B}.insert((hpid_{up/down/left/right}, line))$;
19:        **end if**
20:    **end if**
21: **end for**;
---

## 4 PERFORMANCE OPTIMIZATIONS OF HLC2

To improve the overall performance, the performance modelling to locate performance bottlenecks of the cross-matching using CPU-GPU co-processing is explored to estimate computation overhead, data movement overhead, etc. With an accurate performance model, we profile the cross-matching method in the basic HLC2 and make theoretical analysis to guide further optimization.

### 4.1 Performance model and profiling

#### 4.1.1 Model definition

Our execution time estimation model captures a *host* (CPU) and a *device* (GPU) and executes algorithms in rounds. It is proposed based on ATGPU (Carroll & Wong 2017) model and $\lambda$-Model (Riahi, Savadi & Naghibzadeh 2020), but we capture more accurate and comprehensive estimation. Table 1 summarizes the major notations used in the proposed model. To set up the whole architecture, we let the number of required rounds of the program be represented by $R$ and the set of multiprocessor (MP) be represented by $MP = \{mp_1, mp_2, \ldots, mp_k\}$ with $r$ cores in total and $b$ cores per MP. A round begins with the input data being transferred from host to the global memory of device. Next, the kernel function is run on all or part of MPs and cores. Instructions are executed on cores of MP in the meantime, and the cost of SM executing an instruction is expressed as variable $\gamma$, which corresponds to the clock rate of the GPU. Numbers of global memory blocks accessed by all MPs in the round is represented as $q_i$, with each access cost denoted by $\lambda$. In addition, we denote the maximum number of operations across all MPs as $t_i$. When executing a memory access instruction, the core waits until all kernel memory requests are resolved. Each round ends with the output result being transferred from global memory to the host. Synchronization operations (e.g. de-allocating and reallocating of data structures, resetting the device, etc.) occur with a fixed cost $\sigma$, and the subsequent round starts.

**Table 1.** List of parameters used in the performance model.

| Notation | Description |
|---|---|
| $R$ | the number of required rounds of the program |
| $r$ | the number of cores in total |
| $b$ | the number of cores per MP |
| $\gamma$ | the cost for a MP to execute a single instruction |
| $\lambda$ | the cost to access a memory block in global memory |
| $q_i$ | the number of global memory blocks accessed by all MPs in the round |
| $t_i$ | the maximum number of operations across all MPs executed in the round |
| $\sigma$ | the fixed cost synchronization tasks that need to take place |
| $\alpha$ | the initial latency overhead of sending the first byte between CPU and GPU |
| $\beta$ | the cost required to send each subsequent byte between CPU and GPU |
| $s_i$ | the size of the data transferred between CPU and GPU for $\beta$ calculation |
| $t_s$ | the time of transferring data of size $s_i$ between CPU and GPU |
| $I_i$ | the amount of data transferred inward in the round |
| $O_i$ | the amount of data transferred outward in the round |

The time cost of transferring $d$ bytes of data between CPU and GPU can be calculated as follows:

$$T(d) = \alpha + \beta \times d. \quad (5)$$

The first main value $\alpha$ is a fixed delay for sending the first byte. And $\beta$ represents the time required to send each subsequent byte, which corresponds to the inverse of the transfer bandwidth (Boyer, Meng & Kumaran 2013). According to the method proposed by Riahi et al. (2020), we measure the transfer time of a single byte in the current experimental environment to represent $\alpha$. Then to determine $\beta$, we measure the time $t_s$ of a data transfer of size $s_i$, and then variable $\beta$ can be set to $\frac{t_s - \alpha}{s_i}$. Let $I_i$ ($O_i$ resp.) represents the amount of data transferred inward (outward resp.) in round $i$. So, the cost of inward data transfer can be measured by $T_I(i) = \alpha + \beta I_i$. Likewise, the cost outward data transfer $T_O(i)$ is equal to $\alpha + \beta O_i$. Then assuming that there are sufficient resources in GPU to run each thread block concurrently, the upper limit overhead of the algorithm can be expressed as:

$$\sum_{i=1}^{R} \left( T_I(i) + \frac{t_i + \lambda q_i}{\gamma} + T_O(i) + \sigma \right). \quad (6)$$

*4.1.2 Profiling of cross-matching algorithm*

Based on our proposed model, we can analyse the cost function of cross-matching algorithm, which can be viewed as the upper bound of the cost, and obtain the scaling relation with the data size. The number of rounds is 1. The parallel time complexity is $O(nb)$, the global memory used is $O(n^2)$, the transfer complexity is $O(\alpha + \beta n^2)$, the I/O complexity is $O(\left(\frac{n}{b}\right)^2 (n+b))$, and finally the total cost is

$$O(\alpha + n^2 \frac{t_s - \alpha}{s_i} + \frac{nb + \left(\frac{n}{b}\right)^2 (n+b)\lambda}{\gamma} + \sigma). \quad (7)$$

We assume that two data sets of the same size $n$ is used for calculation. So $n$ entries from one catalogue are compared one by one with the $n$ entries from another catalogue before any indexing and block scheduling optimization are used. Catalogues are then divided into blocks of varying sizes after the index is used, represented by $n' = n'_1, n'_2, \cdots, n'_j$ and $n'' = n''_1, n''_2, \cdots, n''_i$, respectively. Only data blocks of the same index are computed with each other, represented by $v = n'_1 n''_1 + n'_5 n''_5 + \cdots + n'_i n''_i$, which reduces a significant portion of the computation. Therefore, in experiments with real data, the variable $n$ in equation (7) can be assigned as $v$ to calculate the amount of computation for a specific data set. Note that to make the specific cost prediction, the other variables can be determined according to the experimental environment in use. So, it can be seen from the cost function that: (1) as the size of data increases, the data transfer cost increases squared; (2) when the hardware environment is determined, the parallel and I/O complexity have a major impact on the performance, which can be affected by partitioning strategy and parallel memory access patterns.

We conduct pre-experiments with the basic HLC2 presented in Section 3 with real catalogues of different scales from $n = 0$ to 200 MB. Each trial is done 10 times, and the results are averaged to reduce the random error. And we utilize our proposed model to carry out theoretical analysis. Note that the variables related to the specific hardware environment are determined according to the environment used in our experiments (shown in Table 3). The observed results are plotted in Fig. 4. We plot the actual calculation time and the predicted cost by our model in Fig. 5, of which the results are normalized to a scale of 0–1.

*4.1.3 Discussion*

It can be seen from Fig. 5 that our model can accurately predict the execution time under various sizes of input catalogues. Moreover,

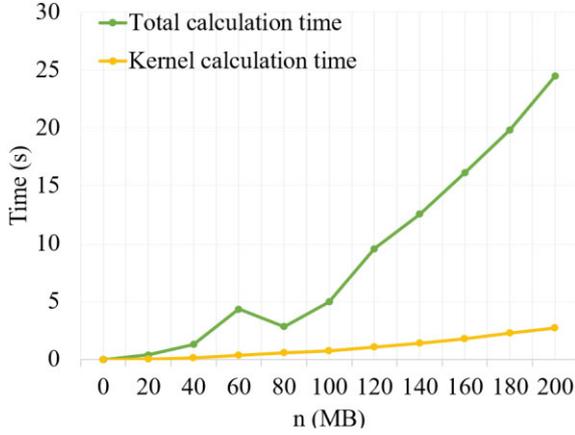

**Figure 4.** Total computation time versus kernel computation time. The hardware configuration used in this preliminary experiment is shown in Fig. 3 and the data set is from public released catalogues. It can be seen that the data transfer takes most of the total time, which has significantly dragged down the overall performance.

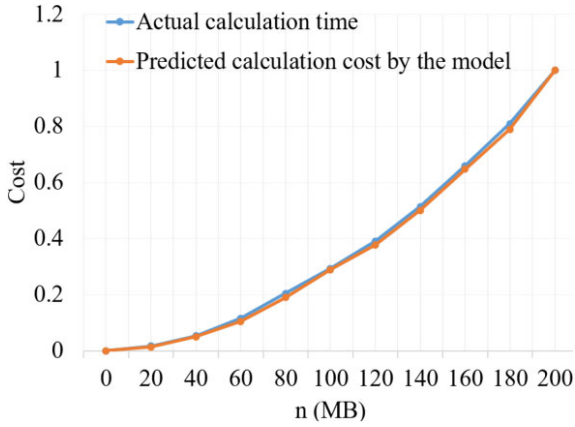

**Figure 5.** Actual total running cost versus predicted cost by our model. The hardware configuration used in this preliminary experiment is shown in Fig. 3 and the data set is from public released catalogues. Results are normalized to a scale of 0–1.

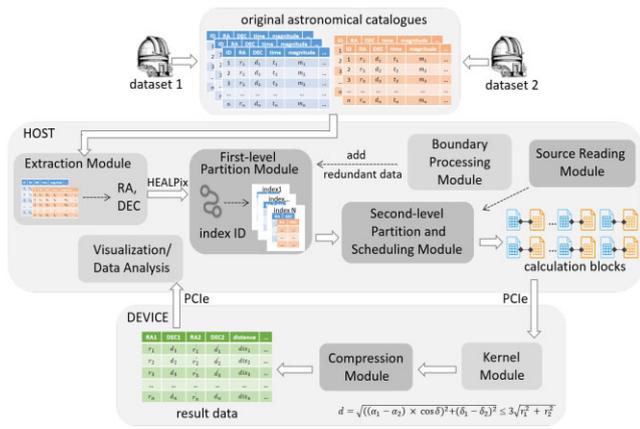

**Figure 6.** Functional extensions of the HLC2. We integrate four new modules for performance optimization based on the basic HLC2 framework. Calculation blocks can be obtained after the two levels of partition for storing and parallel accessing catalogues on the GPU. Source reading module is designed to retrieve the current CPU and GPU computing status to dynamically adjust the splitting strategy. Finally, I/O optimization is implemented on the compression module.

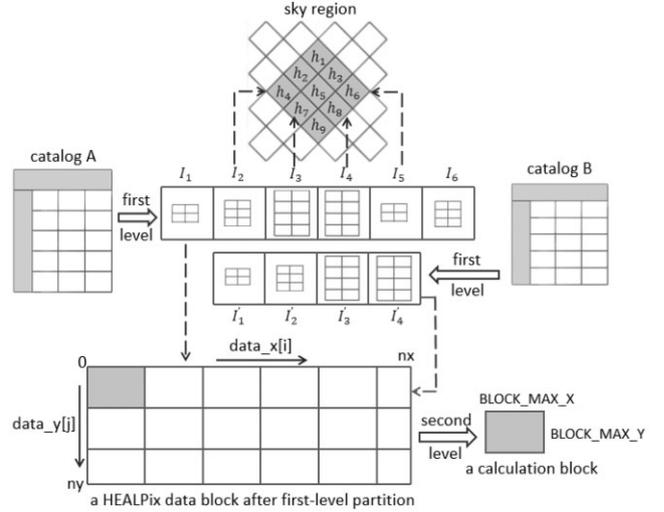

**Figure 7.** Generation process for calculation blocks by two-level partitioning strategy.

as shown in Fig. 4, the total running time rises dramatically and is much steeper than the kernel time. On average, data transfer takes approximately 90 per cent of the total time, meaning that data transfer between CPU and GPU has significantly dragged down the overall performance. It is consistent with the mathematical model. In addition, we find that when catalogue data are densely gathered in a certain sky area, it may exceed the global memory capacity of GPU, so it can not be transferred and calculated. Therefore, combining the bottleneck analysis and existing challenges of cross-matching calculation, HLC2 can be improved to achieve better performance that should be guided by the basic criteria:

(i) On the CPU, an efficient pre-processing scheme is needed to reduce the computational complexity, thereby reducing the data transfer overhead from the source. The topology of the partition method should make most of cross-matching calculation within the block and should be able to support parallel access with high throughput.

(ii) After extracting enough parallelism on the CPU, the input, result, and the spatial distance calculation of calculation blocks should be as independent as possible on the GPU, so as to achieve the overlap of calculation and transmission and maximize parallel kernel computation thus offset part of the I/O overhead.

### 4.2 Two-level partitioning strategy on the CPU

HLC2 is optimized and extended according to the basic criteria obtained from the profiling. As shown in Fig. 6, we integrate four new modules. On the CPU, a two-level partitioning optimization is designed to accurately limit the scope of matching and pre-filter out a lot of useless calculations. Moreover, an efficient data structure named calculation block is built for efficient storing and accessing catalogues. On the GPU, thread organization and I/O optimization strategies are proposed to speed up the algorithm. Fig. 7 demonstrates the computational block generation process using the two-level partitioning strategy. It is worth mentioning that an important concept representing the output of our data partitioning is calculation block. Calculation block is the basic unit of distance calculation of objects in catalogues. Distance calculation only occurs when data from different catalogues are in the same pair of calculation blocks. It is also the basic unit of task allocation in parallel program. Meanwhile, another concept is HEALPix data block, which is generated after the first-level partition. A calculation block is composed of multiple or part of a HEALPix data block. Here, the HEALPix indexing schema is adopted in first-level partition module because of its high efficiency and acceptance among astronomers. This paper extends the original *ang2pix_nest* function of HEALPix library into the *hp_nestid* function, given a partition level (*order*) and position of a celestial body (*RA*, *DEC*), pixel numbers (nested) corresponding to angles theta and phi can be obtained. The pixel number multiplied by a coefficient is regarded as the index of catalogues to convert all indices into the int format. For example, as shown in Fig. 7, the original catalogue input is reconstructed into the map format, HEALPix data blocks are indexed by $\{I_1, I_2, I_3, I_4, I_5, I_6\}$ and $\{I'_1, I'_2, I'_3, I'_4\}$ respectively, each of which may contains different number of celestial records. Assuming that $I_2, I_3, I_4$, and $I_5$ are equal to $I'_1, I'_2, I'_3$, and $I'_4$, respectively, data except these four parts will not be involved in the subsequent calculation. We can conclude that these data are initially distributed in sky regions $h_4, h_6, h_7$, and $h_8$ correspondingly. Thus, workload of searching-related operations is reduced and the distribution of the catalogue data covering the sphere is more reasonable.

As stated in Section 4.1.3, one critical issue of implementing inter-catalogue parallelization is the limited memory size of CPU and GPU. For efficient utilization of resources, proper allocation of memory is also a difficulty in CUDA programming. Therefore, when hardware configuration changes, one-level partition cannot guarantee correct calculation. To address these issues, the source reading module is designed to retrieve the current CPU and GPU computing status. Then, second-level partition and scheduling module dynamically generates the maximum number of catalogue records, which can be computed on CPU (*N*) and GPU (*block_max_x* and *block_max_y*), respectively. Catalogue records that do not exceed the maximum values are extracted and combined to form a calculation block. It is convenient for astronomers to deploy the algorithm on various hardware environments and the computation of corresponding parameters can be adjusted according to various designs of algorithm and data type. For example, the calculation of *N* and *block_max_x* in our algorithm is as follows.

$$N \times \text{sizeof(double)} \times 4 + N^2 \times \text{sizeof(double)} \leqslant \text{sizeof(availableCPUMemory)} \quad (8)$$

$$block\_max\_x \times \text{sizeof(double)} \times 4 + block\_max\_x^2 \times \text{sizeof(double)} \leqslant \text{sizeof(availableGPUGlobalMemory)} \quad (9)$$

The constant '4' in the formula represents the number of variables of right ascension and declination from the two input catalogues. *block_max_y* is calculated in the same way as *block_max_x*.

### 4.3 Optimizations on the GPU

#### 4.3.1 Parallelization hierarchy

As described in Section 4.2, the input original catalogues are divided into calculation blocks and assigned to the GPU. For optimized thread management of every calculation block, we divide the threads in the *x* direction, which means that each thread corresponds to an celestial record in the first data set. Then each thread performs angular distance calculation with every record in *y* direction of the calculation block (second data set) to achieve coalesced access. A two-level thread hierarchy, consisting of thread block and grid is used. Since all of the catalogue records in calculation blocks are stored as 1D array, the thread hierarchy is consisted of 1D grid and 1D block, namely *gridDim.y* = *gridDim.z* = 1 and *blockDim.y* = *blockDim.z* = 1. Given various values of *blockDim.x*, the *gridDim.x* is dynamically calculated by the equation:

$$gridDim.x = (x\_band + blockDim.x - 1)/blockDim.x. \quad (10)$$

Then a kernel with multiple independent threads is launched on the GPU. Threads will follow the same execution path with no warp divergence because all distance calculations are aligned to the same schema. No synchronization within a warp or between thread blocks is necessary. Furthermore, the asynchronous memory transfer is performed by assigning multiple CUDA streams, so that the computation and data transfer can overlap in time.

#### 4.3.2 Memory management

In the GPU global memory, we allocate the following space: the array of position data of celestial objects (*d_in_ra*1, *d_in_dec*1, *d_in_ra*2 and *d_in_dec*2), the amount of catalogue data corresponding to each calculation block (*x_band*, *y_band*), and the cross-matching results of each catalogue pair (*d_out_dis*). We have considered that a critical issue of implementing inter-catalogue parallel distance computation on GPU is the limited size of memory. So, when performing data division before it gets transferred to GPU, the maximum amount of data that can be processed in parallel has been estimated as stated in Section 4.2 to get rid of this problem. Page-locked memory optimization is also implemented to reduce the latency of data access. Meanwhile, we have considered putting part of catalogues into GPU shared memory for fast memory access. Unfortunately, when the catalogue data exceeds 100 MB, shared memory strategy is accompanied by the emergence of frequent data transfer, hence introducing higher transmission delay. Section 4.1 proves that data movement is critical to the performance of GPU accelerated kernels especially, and the catalogues transferred to GPU are only calculated once. So, in HLC2, we did not use the GPU shared memory. Furthermore, the space to store the result needs to be the product of the space of the input arrays from two catalogues, but the angular distance is only stored in the matched position. Instead of the angular distance, we use '1' to mark the location of the matching success, thus reducing the size of the transferred result by half.

## 5 EXPERIMENTS AND RESULTS

In this section, detailed evaluations are performed for HLC2. We evaluate the performance impact of 'HEALPix level' and thread organization strategy. Next, the proposed quad-direction strategy for boundary problems is evaluated. Lastly, overall performance of HLC2 is analyzed comprehensively and compared with other state-of-the-art methods.

All experimental data were initially obtained from public released catalogues from various astronomical projects. In order to show the scalability and applicability of the proposed method, we extracted catalogues with same or different numbers of rows from real surveys and designed D1–D6 data sets (shown in Table 2). For example, data set D3 is generated from a publicly available data set of GAIA by selecting two different numbers of records from the range of RA in [22.5125°, 90.0009°] and DEC in [0.0056°, 9.9999°], while data set D4 is for self-matching scenario. Apart from the real catalogue data, we also make simulated data sets based on the real data for experiments on boundary problems. We have adjusted the data distribution and make it more at the edge of the HEALPix-indexed sky regions. So, it is practical to detect the effect of the boundary processing strategies. Note that the simulated data sets are only used for testing the boundary resolution strategies. In addition, our experiments are conducted on two servers with different CPU and GPU architectures, which are simply referred to as P100 and V100 below. The detailed hardware configurations are listed in Table 3.

**Table 2.** The information of data sets in the experiment of HLC2.

| Name | Sky survey | File size | Description | Number of objects |
|---|---|---|---|---|
| D1 | SDSS × TwoMass | 13.6 M × 40 K | RA ∈ [340.0001°, 360.0000°], DEC ∈ [1.0000°, 1.2000°] | 221 126 × 784 |
| D2 | SDSS × SDSS | 80 M × 80 M | RA ∈ [340.0001°, 360.0000°], DEC ∈ [1.0000°, 1.2000°] | 1 105 627 × 1 105 627 |
| D3 | GAIA × GAIA | 1 G × 400 M | RA ∈ [22.5125°, 90.0009°], DEC ∈ [0.0056°, 9.9999°] | 12 935 566 × 5 420 000 |
| D4 | TwoMass × TwoMass | 3.7 G × 3.7 G | RA ∈ [0.0000°, 360.0000°], DEC ∈ [−89.9928°, 89.9956°] | 61 734 000 × 61 734 000 |
| D5 | TwoMass × TwoMass | 9 G × 1.5 G | RA ∈ [0.0000°, 360.0000°], DEC ∈ [−89.9928°, 89.9956°] | 144 046 000 × 25 722 500 |
| D6 | TwoMass × TwoMass | 28.5 G × 6.2 G | RA ∈ [0.0000°, 360.0000°], DEC ∈ [−89.9928°, 89.9956°] | 467 555 800 × 102 890 000 |

**Table 3.** Hardware configurations for experiments.

| Model | Processor | Architecture | Clock (GHz) | Memory size (GB) | Number of SM | Cores | CUDA capability | Compiler |
|---|---|---|---|---|---|---|---|---|
| Intel Xeon E5-2682 | CPU | BroadWell | 2.5 | 60 | - | 32 | - | gcc 7.5 |
| NVIDIA Tesla P100 | GPU | Pascal | 0.715 | 16 | 56 | 3584 | 6.0 | nvcc 10.2 |
| Intel Xeon Silver4114 | CPU | Skylake-EP | 2.2 | 32 | - | 16 | - | gcc 7.5 |
| NVIDIA Tesla V100 | GPU | Volta | 1.246 | 16 | 80 | 5120 | 7.0 | nvcc 10.2 |

The data samples and source codes of HLC2 are published at https://github.com/Melody888Evan/HLC2.

### 5.1 Impact of HEALPix level

To ensure appropriate load balancing after the two-level partitioning is performed, it is necessary to avoid clustering large proportion data into just a few HEALPix data blocks. Thus, to select moderate 'HEALPix level', we measure the balance of completeness and performance between different choices. Data sets D2, D3, and D5 are used and indices based on 'HEALPix level' are generated varied from 5 to 13. A default setting of 64 threads per block is used. Fig. 10 illustrates the results of the total calculation time and the number of matched sources as the granularity of the partitioning varies. Since the number of matches varies for data sets of different sizes in the experiments, we use the ratio[1] of each result to the maximum result for each data set to represent the absolute number of matches to better show the trend of matched results with the partitioning levels. It can be seen that for the three data sets, the behaviour of completeness of decreases with increasing levels of partitioning. It is consistent with the fact that finer granularity of partitioning increases the possibility of matching loss. As for execution time, D2 reaches a low point in level 10. D3 and D5 achieve the peak performance in levels 8 and 9, respectively. For D3, setting of level 8 reports up to 12.1 times faster than level 13 and increases 1.4 per cent from level 13 to level 8. Because the celestial record is only computed with the record in the same block, small partitions will reduce computational complexity to some extent. But when the data volume rises to 1G, overly detailed partition brings high consumption to data pre-processing and the generation of redundant data. As a result, total cross-matching time is expected to continue increasing after the best point of index level.

---

[1] $x^* = \frac{x}{x_{max}}$, $x^*$ represents the proportional results used in the plot, $x$ represents each of the original data, and $x_{max}$ refers to the maximum result value under the input of different partition levels. Since it takes a long time to obtain the results without match loss for the large catalogues, we use the ratio of each result to $x_{max}$ to represent the number of matches.

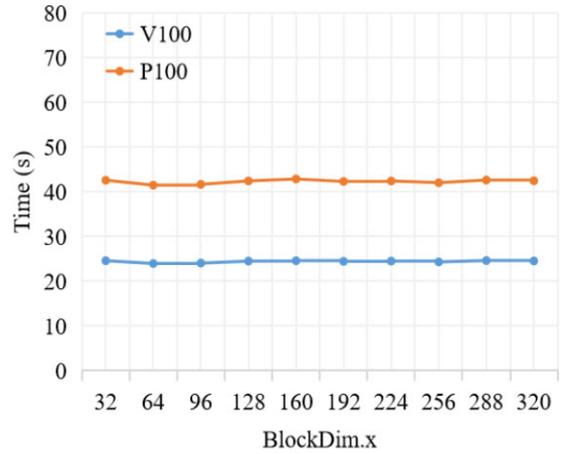

**Figure 8.** Performance with thread organization varied. With data set D3 and HEALPix level 9, the curves show the HLC2 processing time on P100 and V100.

Based on our experimental results, usually it is appropriate to set 'HEALPix level' as 10, 8, and 9.

### 5.2 Impact of thread organization strategy

The impact of thread organization strategy is evaluated on two different GPU environments, with data set D3 and fixed HEALPix level 9. According to the calculation strategy described in Section 4.3, the number of blocks in one grid can be dynamically obtained according to the number of threads in each block. As for server P100 with a CUDA capability of 6.0, the number of SPs is 3584, which is smaller than V100's 5120. Partially due to this reason, the overall cross-matching time of HLC2 on P100 is longer than that of V100. Since there are both 64 SPs in each SM of V100 and P100, one SM can only execute 64 threads at the same time. Therefore, it can be seen in Fig. 8 that the peak performance can be obtained for *BlockDim.x* = 64 on both V100 and P100. In this configuration, maximum thread parallelization can be achieved for each SM with minimal thread scheduling overhead. However, GPU thread organization parameters may also be affected by various other factors like memory bandwidth and GPU clock frequency, so it should be adjusted according to the specific hardware configurations to obtain optimal results. In addition, experimental results show that the I/O problem of the GPU also limits the utilization of computing resources and blocks the further improvement of performance on two servers, which needs to be further solved.

**Table 4.** Comparison of matched record number of indexing schemes using simulated data sets. With the increase of data set size, the boundary problem usually becomes more serious. Our quad-direction strategy is the most accurate compared with other indices.

| Data size | No Index | HTM | HEALPix | Dual Index | Quad-direction Strategy |
|---|---|---|---|---|---|
| 10 M | 665 039 | 662 539 (−2500) | 663 914 (−1125) | 665 039 (−0) | 665 039 (−0) |
| 50 M | 14 366 740 | 14 337 115 (−29 625) | 14 366 740 (−0) | 14 366 740 (−0) | 14 366 740 (−0) |
| 100 M | 58 553 930 | 58 548 430 (−5500) | 58 553 680 (−250) | 58 553 712 (−218) | 58 553 798 (−132) |
| 200 M | 181 467 425 | 181 449 425 (−18 000) | 181 465 925 (−1500) | 181 466 248 (−1177) | 181 466 947 (−478) |
| 500 M | 680 502 844 | 680 452 444 (−50 400) | 680 501 594 (−1250) | 680 501 829 (−1015) | 680 502 079 (−765) |

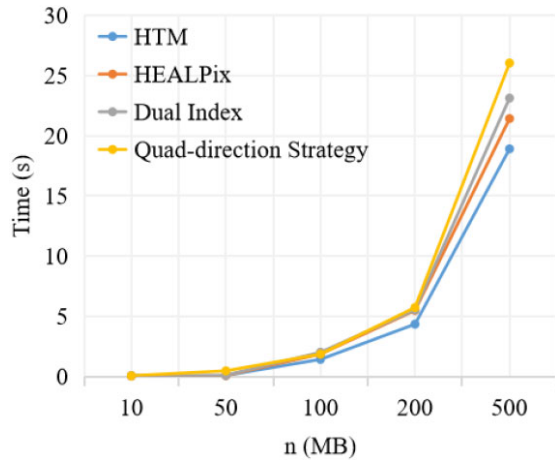

**Figure 9.** Performance with indices varied. With simulated data sets of which data are more distributed at the edge of the indexed sky regions, $BlockDim.x = 64$, the curves show the HLC2 processing time on V100. Our quad-direction strategy shows comparable speed to the next best approach with improvement on completeness illustrated in Table 4.

### 5.3 Evaluation of the quad-direction strategy

The completeness of cross-matching is compared using various indexing schemes on simulated data sets of different sizes with $BlockDim.x = 64$ on V100. The number of records successfully matched and the cross-matching time required by different indexing schemes are illustrated in Table 4 and Fig. 9, respectively. The results in brackets in Table 4 represent the number of results that differ from the accurate serial results without indices. We do not plot the execution time when no index is used, because serial calculation leads to several days of time. It can be seen that when data volume is less than 50M, the total calculation time based on these indices is close, and HTM brings more boundary leakage (0.4 per cent), while the completeness of our quad-direction strategy and dual index are comparable. With the increase of data volume, time of using HTM is the least, but the error brought by HTM has a great influence on astronomical discovery. Compared with dual index, execution with quad-direction strategy takes about the same amount of time, but reduces data loss by about half. So combined with performance and completeness, quad-direction strategy becomes the best among these four indices.

### 5.4 Overall performance evaluation and analysis

As an indication of how HLC2 scales with data size, Fig. 11 shows the running time of HLC2 on data sets D1–D6, illustrated in Table 2. It can be seen that generally the more objects each catalogue contains, the more time it would cost, which shows our method scales well on catalogues of various data volumes. Meanwhile, the relative distribution of objects in the data set also affects the complexity of cross-matching. The best speed-up is not achieved when there are a few objects in the catalogue, because the consumption of I/O and initialization make up a considerable proportion. As the number of objects increases, the matching computation dominates most of the time, which, in turn, reports a better speed-up. Our experiments show that, compared with the calculation results presented in Fig. 4, the data movement time after optimization is reduced by 38 per cent. Furthermore, the performance of HLC2 on P100 and V100 shows similar trends, which confirms that two-level partitioning strategy can be well adapted to different environments.

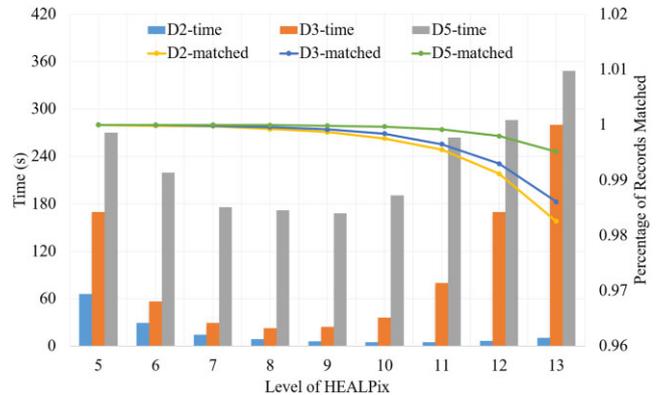

**Figure 10.** Performance with HEALPix level varied. With data sets D2, D3, and D5, and $BlockDim.x = 64$ on V100, the HLC2 processing time and completeness changes with different HEALPix levels. The percentage of each result to the maximum result for each data set is used to represent the number of matched catalogue records.

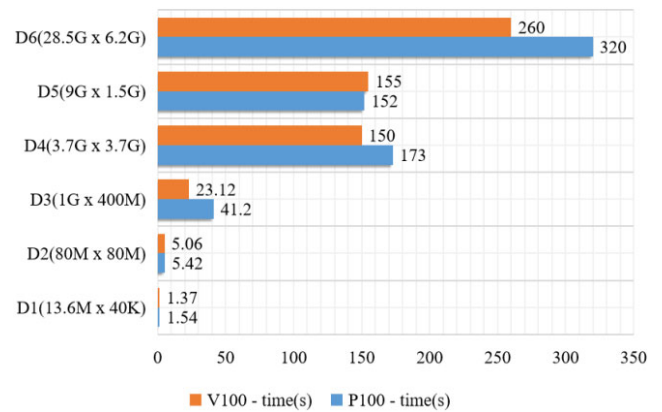

**Figure 11.** Performance with size of catalogues varied. It can be seen that HLC2 can efficiently obtain the results of cross-matching on data set of different sizes, and it can complete the calculation of two hundred-million-scale catalogues in minutes.

**Table 5.** Performance comparison of cross-matchers. It can be seen that our HLC2 has higher processing speed in various sizes of catalogues, especially in large data sets.

| Cross-matcher | Time for small number of objects | Time for medium number of objects | Time for large number of objects |
| --- | --- | --- | --- |
| Zhao et al. (2009) | – | – | 32 min (470 992 970 × 100 106 811) |
| Du et al. (2014) | – | 7 min (946 464 × 470 992 970) | 23 min (470 992 970 × 100 106 811) |
| catsHTM (Soumagnac & Ofek 2018) | – | 160 s (55 395 532 × 55 395 532) | 53 min (470 992 608 × 563 908 224) |
| C3 (Riccio et al. 2017) | 90 s (1 000 000 × 1 000 000) | 800 s (1 000 000 × 10 000 000) | – |
| HLC2 | 5 s (1 105 627 × 1 105 627) | 150 s (61 734 000 × 61 734 000) | 260 s (467 555 800 × 102 890 000) |

To compare HLC2 with other state-of-the-art cross-matchers, we select mainly four representative cross-matchers based on spherical position deviation, namely Zhao et al.'s algorithm, Du et al.'s algorithm, and C3. Algorithms based on other cross-matching algorithms like catsHTM and Jia et al.'s algorithm are also discussed for comparison. They are all start-of-the-art cross-matching algorithms running on multi-CPU and GPU processors. Among these competitors, Du et al.'s algorithm and Zhao et al.'s algorithm are compared in terms of both performance and completeness, and the others are tested by only end-to-end execution time because they did not deal with boundary problems.

The performance comparison of HLC2 and other state-of-the-art cross-matchers is shown in Table 5. Experimental results of HLC2 are compared with the results reported in their papers. Catalogues are divided into small, medium, and large groups by size, and the HLC2 is tested on all sizes. HLC2 on large catalogues achieves speed-ups of more than 7.4 times over Zhao et al.'s bit-operation algorithm, and more than 28 and 5.3 times over Du et al.'s algorithm on medium and large data set, respectively without relatively losing the completeness. The mcatCS only focuses on the matching of one record in the catalogue (e.g. $n$ objects) that is of interest to astronomers, of which the complexity is only $O(n)$. So HLC2 processes much faster than mcatCS. In comparison with C3, HLC2 is around 18 and 2033 faster on small and medium catalogues and has higher adaptability to various sizes of data. For cone-search-based cross-matching like catsHTM, HLC2 is slightly faster on medium sizes and shows greater performance as the amount of data grows. These performance improvements demonstrate the effectiveness of our partition and parallelization optimizations. Compared with our method of $O(mn)$, Jia et al.'s algorithm is based on cone search, which is used in different astronomical scenarios and does not give an end-to-end time, so we exclude it from the table.

## 6 CONCLUSION AND FUTURE WORK

The substantial increasing amount of astronomical observation data establishes the need for efficient cross-matching methods and tools. This paper proposes an efficient and scalable cross-matching framework named HLC2 on CPU-GPU heterogeneous computing environments. Users only need to input two catalogues to be matched, and the cross-matching result can be generated automatically. Additionally, HLC2 provides flexibility for users to perform cross-matching on various architectures of NVIDIA GPU and sizes of catalogues. It can be directly applied to the fusion of large astronomical catalogues from survey missions and astronomical data centres and can perform cross-matching of catalogues with hundred millions of sources in minutes.

Furthermore, we extend the functions and optimize the performance based on the basic framework. A two-level partitioning strategy for the generation of calculation blocks and a quad-direction strategy for boundary problems are proposed on the CPU. The cross-matching processing is also accelerated using GPU through dynamic thread organization and I/O optimization. Experimental results show that HLC2 outperforms the state-of-the-art methods on the overall execution time, especially for large catalogues with millions of celestial records. And the novel quad-direction strategy reduces the result loss by about half using comparable calculation time. The HEALPix level, thread organization, and the distribution of celestial objects may have substantial impact on performance.

In future research, we will continue to explore the boundary problems in large-scale catalogue cross-matching to return complete matches for all sources in an acceptable amount of time. Meanwhile, we will extend HLC2 to clusters of multiple computers with CPUs and GPUs, which can provide more computing power. In that case, a new level of parallelism needs to be considered. With HLC2 as part of core implementation, task transfer, load balancing, and I/O issues with multiple physical nodes are further challenged and worth exploring. HLC2 is also expected to be integrated into the China-VO platform and provides reliable cross-matching services.

## ACKNOWLEDGEMENTS

This work is supported by the National Key Research and Development Program of China (2022YFF0711502), and National Natural Science Foundation of China (NSFC) (12273025, 12133010). Data resources are supported by China National Astronomical Data Center (NADC), Chinese Academy of Sciences (CAS) Astronomical Data Center and Chinese Virtual Observatory (China-VO).

## DATA AVAILABILITY

The codes and data samples are available from https://github.com/Melody888Evan/HLC2. The data used for the analysis in this study are available from the corresponding author on request.

This paper has been typeset from a TeX/LaTeX file prepared by the author.